
\def\nolabels{\def\eqnlabel##1{}\def\eqlabel##1{}\def\reflabel##1{}}
\def\writelabels{\def\eqnlabel##1{%
{\escapechar=` \hfill\rlap{\hskip.09in\string##1}}}%
\def\eqlabel##1{{\escapechar=` \rlap{\hskip.09in\string##1}}}%
\def\reflabel##1{\noexpand\llap{\string\string\string##1\hskip.31in}}}
\nolabels
\global\newcount\meqno \global\meqno=1
\global\meqno=1
\def\eqnn#1{\xdef #1{(\the\meqno)}%
\global\advance\meqno by1\eqnlabel#1}
\def\eqna#1{\xdef #1##1{\hbox{$(\the\meqno##1)$}}%
\global\advance\meqno by1\eqnlabel{#1$\{\}$}}
\def\eqn#1#2{\xdef #1{(\the\meqno)}\global\advance\meqno by1%
$$#2\eqno#1\eqlabel#1$$}
\overfullrule=0pt
\magnification=\magstep1
\font\twelvebf=cmbx12
\nopagenumbers
\footline={\ifnum\pageno>1\hfil\folio\hfil\else\hfil\fi}
\line{\hfil RU-92-17-B}
\line{\hfil CU-TP-584}
\line{\hfil CERN-TH 6768/93}
\line{\hfil January 1993}
\vglue .7in
\centerline {\twelvebf On the Electromagnetic Interactions of Anyons}
\vskip .4in
\centerline{\it Chihong Chou}
\vskip .1in
\centerline { Physics Department, Rockefeller University}
\centerline {New York, New York 10021, U.S.A.}
\vskip .2in
\centerline{\it V.P.~Nair}
\vskip .1in
\centerline{Physics Department, Columbia University}
\centerline{New York, New York 10027, U.S.A.}
\vskip .2in
\centerline{\it and}
\vskip .2in
\centerline {\it Alexios P.~Polychronakos}
\vskip .1in
\centerline{Theory Group, CERN}
\centerline{CH-1211, Gen\`eve 23}
\centerline {Switzerland}
\vskip .5in
\baselineskip=14pt
\centerline {\bf Abstract}
\vskip .1in
Using the appropriate representation of the Poincar\'e group and a
definition of minimal coupling,
we discuss some aspects of the electromagnetic interactions of charged
anyons.
In a nonrelativistic
expansion, we derive a Schr\"odinger-type equation for the anyon wave
function which includes spin-magnetic field and spin-orbit couplings.
In particular, the
gyromagnetic ratio for charged anyons is shown to be 2; this last
result is essentially a reflection of the fact that the spin is parallel
to the momentum in (2+1) dimensions.
\vskip .6in
\noindent
This work was supported in part by the US Department of Energy.
\vfill\eject
\baselineskip 22pt
\def\e{\epsilon}
\def\ho{Heisenberg equations of motion
\eqn\Ho{\eqalign{&{dx^a\over d\tau}={1\over i}[x^a, \ G_0], \cr
		&{dp^a\over d\tau}={1\over i}[p^a, \ G_0]}}}
\def\h\eqn\H{\eqalign{&{dx^a\over d\tau}={1\over i}[x^a, \ G], \cr
		&{dp^a\over d\tau}={1\over i}[p^a, \ G]}}

In two spatial dimensions, it has been known for some time, that
one has the intriguing possibility of anyons or particles of arbitrary
spin and statistics $^1$. They may also be important in some physical
situations. For example, the quasiparticles relevant to the understanding
of fractional quantum Hall effect are considered to be anyons $^2$. In this
paper, we study some aspects of the electromagnetic interactions
of anyons, in particular the gyromagnetic ratio and spin-orbit
couplings.

There are many ways of constructing theories containing anyonic excitations.
One of the most popular ways is to start with a theory of bosons or
fermions and then couple these particles minimally to a U(1) gauge
field, the so-called statistical field, whose dynamics is governed
by the Chern-Simons action. The elimination of the gauge field then
leads to a redefinition of one-particle states. The spin of the redefined
one-particle states is not necessarily an integer or half-integer;
it is a value determined by the coupling to the gauge field $^3$. Simple
as this procedure is, it is not the most economical or minimal way
to understand anyons. As in discussions of point-particles
in (3+1)-dimensional quantum mechanics, one can define
one-anyon states as a unitary representation of the relevant
operator algebra, viz. the (2+1)-dimensional Poincar\'e algebra $^{4,5}$.
(This is constructed as an induced representation $^6$.) One can also
introduce a manifestly covariant wave equation for anyons,
which is the anyonic analogue of the Dirac equation of spin-${1\over 2}$
particles $^{6,7}$.

Since relativistic motion is unimportant in most condensed matter
contexts, one may ask to what extent Poincar\'e or Lorentz symmetry,
rather than Galilean symmetry, is useful. For the Dirac theory, the
answer is well known. If one couples an external electromagnetic
field minimally, one gets, in the nonrelativistic approximation, an
improved Schr$\ddot{\rm o}$dinger equation with a spin-magnetic field coupling
(due to spin magnetic moment) and a corresponding spin-orbit
coupling. In particular the gyromagnetic ratio is 2. In this paper,
we shall not use the covariant wave equation mentioned above.
Instead we shall use the canonical or symplectic framework for the
induced representation of Poincar\'e group appropriate to anyons
$^{5,6}$.
We give a definition of minimal coupling of the electromagnetic
field to anyons in the language of symplectic or canonical
structure and then obtain the improved Schr$\ddot{\rm o}$dinger equation.
In particular, we show that the gyromagnetic ratio $g$ is 2 for anyons
as well and there are the corresponding spin-magnetic field and spin-orbit
couplings. We emphasize that these results only depend on symmetry
and our definition of minimal coupling. They are universal,
independent of any model for the realization of anyonic
excitations. In (2+1) dimensions, the spin is parallel to the momentum,
and as we argue below (after equation (19)), this is really all we need
to show that $g=2$.

We shall begin with a discussion of a spinless charged particle
in an electromagnetic field. A spinless particle may be described
by a set of momentum variables $p^a$ and position
variables $x^a$, $a=0,1,2$. The canonical structure or symplectic
two-form is given by
\eqn\a{\eqalign{\omega_0&\equiv {\textstyle{1\over 2}} \omega_{0, AB}\ d\xi^A
       \wedge d\xi^B \cr
       &=dx^a \wedge dp_a,}}
where $\xi^A=(p^a, x^a)$. The Poisson brackets are given
by $\{\xi^A,\ \xi^B\}=(\omega_0^{-1})^{AB}$. (Thus the symplectic
two-form may be regarded, albeit somewhat crudely, as a
succinct way of specifying Poisson brackets.) The introduction
of electromagnetic field $A_a$ by the minimal prescription
$p_a\rightarrow p_a-eA_a$ is equivalent to $\omega_0\rightarrow
\Omega_0=\omega_0+eF$, where $F={1\over 2} F_{ab}dx^a\wedge
dx^b$, $F_{ab}=\partial_a A_b-\partial_b A_a$
is the
electromagnetic field strength and $e$ is the charge of the
particle.

The motion of the charged particle is given by the (classical)
Lorentz equations, viz.
\eqn\2{\eqalign{&{p^a\over m}={dx^a\over d\tau}, \cr
	        &{dp^a\over d\tau}=-{e\over m}F^{ab}p_b,}}
$\tau$ is the parameter for the trajectory of the particle
(with mass $m$). We have chosen a specific parametrization
or equivalently chosen a gauge-fixing for the gauge freedom
of reparametrizations of the trajectory and so the equations
\2 are not invariant under reparametrizations. Equations \2 tell
us that the infinitesimal change of $\tau$ is given, on
the phase space, by a vector field
\eqn\3{V={p^a\over m}{\partial \over \partial x^a}-
	{e\over m}F^{ab}p_b{\partial\over \partial p^a}.}
The canonical generator of $\tau$-evolution, say $G_0$, is
defined by $^8$
\eqn\4{V\rfloor\Omega_0=-dG_0.}
($\rfloor$ denotes interior contraction, i.e. $v\rfloor \sigma =
v^a \sigma_{ab}dx^b$, for $v=v^a \partial_a$ and $\sigma = {1\over 2}
\sigma_{ab}dx^a\wedge dx^b$.)
This gives $G_0=-p^2/(2m)+constant$. Anticipating the eventual
value of the constant, we choose it to be $m/2$. (In a
sense, this will be the definition of the mass.) Thus
\eqn\5{G_0=-{1\over 2m}(p^2-m^2).}

Since we need reparametrization invariance, $\tau$-evolution must be
trivial. Thus we must set $G_0\approx 0$. In the quantum theory we
write
\eqn\6{G_0 \psi =0.}
 From $\Omega_0=\omega_0+eF={1\over 2}\Omega_{0,AB}~d\xi^A\wedge
d\xi^B$, we have
\eqn\7{\Omega_{0,AB}=\left(\matrix{0& -g_{ab}\cr g_{ab}& eF_{ab}\cr}
		     \right),}
where the metric $g_{ab}=diag(+--)$. This gives the
commutation rules
\eqn\8{\eqalign{&[x^a, \ x^b]=0, \cr
	        &[p^a, \ x^b]=ig^{ab}, \cr
		&[p^a, \ p^b]=ieF^{ab}.}}
These have the solutions
\eqn\9{p_a=i\partial_a + e A_a,}
and equation \6 becomes
\eqn\a{[(\partial_a-ieA_a)^2-m^2]\psi=0.}
This is the Schr$\ddot{\rm o}$dinger equation (in this case, the Klein-Gordon
equation) which we are seeking.
One can easily show that the Lorentz equations \2 \ are
quantum mechanically realized through the \ho \ with the
commutation rules \8.

The strategy of obtaining the Schr$\ddot{\rm o}$dinger-type equation is now
clear. We start with the symplectic two-form. From the Lorentz equations
we find  the generator of $\tau$-evolution. Setting this generator to
zero on the wave functions gives us the equation we are seeking. To
realize this as a differential equation we solve the commutation
rules in terms of a set of coordinates and their derivatives (namely,
canonical variables or Darboux coordinates).

For a free anyon with spin $-s$, the symplectic structure is given by
$^{5,6}$
\eqn\b{\omega=dx^a\wedge dp_a + {\textstyle{1\over 2}}s\e_{abc}{p^a dp^b
\wedge dp^c
\over (p^2)^{3/2}}.}
The commutation rules are given by
\eqn\c{\eqalign{&[x^a, \ x^b]=is~\e^{abc}{p_c\over (p^2)^{3/2}}, \cr
	        &[p^a, \ x^b]=ig^{ab}, \cr
		&[p^a, \ p^b]=0.}}
Consider the Lorentz generator $J^a$ defined by
\eqn\d{\eqalign{&[J^a, \ p^b]=i\e^{abc}p_c, \cr
	        &[J^a, \ x^b]=i\e^{abc}x_c.}}
It is easy to see that $J^a$ is given by
\eqn\f{J^a=-\e^{abc}x_bp_c-s{p^a\over \sqrt{p^2}}.}
We can solve the commutation rules \c \ in terms of canonical
variables as
\eqn\g{\eqalign{x^a&=q^a+\alpha^a(p) \cr
		&\alpha^a(p)=s\e^{abc}{p_b \eta_c
		\over p^2+\sqrt{p^2}p\cdot\eta},}}
where $\eta^a=(1,0,0)$ and $[q^a,q^b]=0,~[p^a,p^b]=0,~
[p^a, q^b]=ig^{ab}$ or $q^a=-i{\partial
\over \partial p_a}$.

Using equation \g \ for $x^a$ in equation \f, we can write
\eqn\h{J^a=-i\e^{abc}p_b{\partial\over\partial p^c}-s
	{p^a+ \sqrt{p^2}\eta^a \over \sqrt{p^2}+p\cdot\eta}.}
We see that $p\cdot J+s\sqrt{p^2}=0$. With $p^2=m^2$, we see
that the spin is indeed $-s$. Thus the symplectic structure $\omega$
of equation \b \ is indeed appropriate to anyon.

The symplectic structure $\Omega$ for anyons in an electromagnetic
field is now obtained by $\omega\rightarrow\Omega=\omega+eF$. Thus
\eqn\i{\Omega=dx^a\wedge dp_a+ {\textstyle{1\over 2}}s \e_{abc}{p^a dp^b
\wedge dp^c
\over (p^2)^{3/2}}+ {\textstyle{1\over 2}}eF_{ab}dx^a\wedge dx^b
+~{\cal O}(\partial F).}
With the introduction of spin it is possible that $\Omega$ has
further corrections depending on the gradients of the field
strength $F$. This is indicated by ${\cal O}(\partial F)$ in equation \i.
To the approximation that we are interested in, these terms
will not be important. Likewise, we take the equations of motion
to be
\eqn\j{\eqalign{&{dx^a\over d\tau}={p^a\over m}, \cr
	        &{dp^a\over d\tau}=-{e\over m}F^{ab}p_b
		+~{\cal O}(\partial F),}}
i.e. we have the Lorentz equations, again upto possible terms
which depend on gradients of $F$. We can now determine the
vector field of $\tau$-evolution and proceed to obtain a wave equation.
However, before doing so, we point out that already one can see that the
gyromagnetic ratio must be 2. For this, consider the precession
equation for the spin vector $S^a$
in an external electromagnetic field. It is given by the
Bargmann-Michel-Telegdi (BMT) equation$^9$

\eqn\k {{dS^a\over d\tau} = -{eg\over 2m} \left[ F^{ab}S_b -{p^a (p_k
F^{kl}S_l)\over p^2}\right] ~- {p^a\over p^2} (S_k{dp^k\over d\tau})
+{\cal O}(F^2, \partial F)}
The BMT-equation only requires the constancy of $p\cdot S$, at least
to the order in $(F, \partial F)$ that we are interested in; thus it
is true for anyons as well. If one uses the equations of motion (19),
we get
\eqn\l{{dS^a\over d\tau}= -{eg\over 2m}F^{ab}S_b ~+~ {e\over 2m}(g-2)
{p^a (p_k F^{kl}S_l)\over p^2}+{\cal O}(F^2, \partial F)}
For a free anyon, the spin vector is parallel to the momentum;
indeed, from (15), we have $S^a =-{sp^a\over {\sqrt {p^2}}}$.
With electromagnetic fields, we expect
\eqn\m{ S^a =-{sp^a\over {\sqrt {p^2}}} ~+~ {\cal O} (F,\partial F)}
Using the Lorentz equations (19), we thus get
\eqn\n{ {dS^a\over d\tau} = -{e\over m}F^{ab}S_b~+~ {\cal O}(F^2,
\partial F)}
Comparing with equation (21), we see that we must have $g=2$. The simple
physical reason for this result is that, in (2+1) dimensions, spin and
momentum are parallel, with corrections perhaps of order $F$.

We now turn to the wave equation. The vector field of
$\tau$-evolution is given by $V$ of equation \3 \ with possible
${\cal O}(\partial F)$ corrections. The generator of $\tau$-evolution, $G$,
defined by $-dG=V\rfloor\Omega$, is identified from equation \i \ as
\eqn\o{G=-{1\over 2m}(p^2+2es{F\cdot p\over \sqrt{p^2}}-m^2),}
where $F_{ab}=\e_{abc}F^c$. In this calculation we have neglected
the gradients of $F$; otherwise the vector field $V$ is not
Hamiltonian, i.e. $V\rfloor\Omega$ is not  the derivative
of some function $G$. More generally one can introduce corrections
to $V$ so that it becomes Hamiltonian; in general one has to modify
$\Omega$ as well. Although we shall not pursue this, in fact, one
can do a systematic expansion in powers of $\partial F$ and higher
derivatives, zigzagging between corrections in $V$ and corrections
in $\Omega$. Notice however that our expression for $G$ is correct to
the first order in the spin $s$ even for a nonuniform field. In this
case, the equation for $p^a$ has a correction term $(-es/m) {p^b\over
{\sqrt {p^2}}}\partial^a F_b$, which is precisely the extra force on a
dipole of magnetic moment $\mu =es/m$, which corresponds again to a
gyromagnetic ratio 2. In fact, one can turn this argument around by
introducing an arbitrary magnetic moment and the corresponding force in
a nonuniform field; then the requirement that the corresponding vector field
$V$ be Hamiltonian will lead us to $g=2$.

The wave functions obey the equation $G\psi=0+~{\cal O}(\partial F)$,
namely,
\eqn\p{(p^2+2es{F\cdot p\over \sqrt{p^2}}-m^2)\psi=0.}
We need to simplify the commutation rules in order to write
this as a differential equation. $\Omega$ gives the commutation
rules as
\eqn\q{\eqalign{&[x^a, \ x^b]=i({\tilde M}^{-1}f)^{ab}, \cr
	        &[p^a, \ x^b]=i(M^{-1})^{ab}, \cr
		&[p^a, \ p^b]=i(M^{-1}F)^{ab},}}
where $f_{ab}=s\e_{abc}p^c/(p^2)^{3/2}$ and $M_{ab}=g_{ab}+
e(Ff)_{ab}~\equiv g_{ab}+e F_{ak}f^k_{~b}$. $\tilde M$ denotes the
transpose of $M$.

It is, of course, very difficult to solve these equations in general.
We shall solve \q \ in terms of a series in powers of $F$. To the
lowest nontrivial order, which is all we need for the discussion
of magnetic moments and spin-orbit couplings, equation \q \ becomes
\eqn\r{\eqalign{&[x^a, \ x^b]=if^{ab}-i(fFf)^{ab}, \cr
	        &[p^a, \ x^b]=ig^{ab}-i(Ff)^{ab}, \cr
		&[p^a, \ p^b]=iF^{ab}.}}
Notice that, to the order we are considering, there is no ordering
problem in the commutation rules. We shall now solve these in terms
of coordinates $q^a$ and momenta $k^a$ as follows.
\eqn\s{\eqalign{&p_a=k_a-{\textstyle{1\over 2}}eF_{ab}[q^b+2\alpha^b(k)]+
{}~{\cal O}(F^2,\partial F, \cdots), \cr
&x_a=q_a+\alpha_a(k)+{\textstyle{1\over 2}}eF_{bc}\left({\partial \alpha_a\over
\partial k_c}q^b +{\partial\alpha^c\over \partial k^a}\alpha^b\right)
+~{\cal O}(F^2,\partial F, \cdots),}}
where
\eqn\t{[q_a, \ q_b]=0, \ \ [k_a, \ k_b]=0, \ \ [k_a, \ q_b]=ig_{ab},}
and $\alpha^a(k)$ is given by equation \g. Notice that $x_a$ and $p_a$
in equation \s \ are clearly hermitian operators. One can also
verify that the Lorentz equations \j \ are realized in the quantum
theory as Heisenberg equations of motion by using \o , \r \  and \s.

The quantity $-{1\over 2}F_{ab}q^b$
is the electromagnetic vector potential $A_a$ for a constant strength $F$,
or a slowly varying field in the sense of our approximation.
Using equations \s, \t, the constraint of reparametrization invariance,
viz. equation \p \ becomes
\eqn\u{\left[(k_a+eA_a-eF_{ab}\alpha^b)^2+
2es{F\cdot k\over \sqrt{k^2}}-m^2\right]\psi=
0 + ~{\cal O}(F^2, \partial F, \cdots).}
We can further simplify this equation as
\eqn\v{\left[(k_a+eA_a)^2+ 2e\e_{abc}k^aF^b\alpha^c+
2es{F\cdot k\over \sqrt{k^2}}-m^2\right]\psi=
0 + ~{\cal O}(F^2, \partial F, \cdots).}
We can now obtain the nonrelativistic equation by writing the
wave function $\psi$ in terms of the nonrelativistic wave function
$\psi_{\rm NR}$ as
\eqn\w{\psi=e^{-imq^0}\psi_{\rm NR}.}
This gives $k_0=m+H$ where $H$ is the operator $i{\partial\over\partial
q^0}$ on $\psi_{\rm NR}$. One can take $H$ to be small compared to $m$
and expand \v \ to first order in $H$. We then find
\eqn\x{H\psi_{\rm NR}=
\left[{({\bf k}+e{\bf A})^2\over 2m}-{e\over m}\e_{abc}k^aF^b\alpha^c
-{es\over m}{k_a F^a \over\sqrt{k^2}}~+~eA_0\right]\psi_{\rm NR}+~{\cal
O}(F^2,\partial F, {1\over m^2},\cdots ).}
Further, we see that in our approximation
\eqn\y{\eqalign{{k_aF^a\over \sqrt{k^2}}&\approx F_0-
		{{\bf k}\cdot {\bf F}\over m}, \cr
		\e_{abc}k^aF^b\alpha^c &\approx
		{s\over 2m}{\bf k\cdot F}.}}
Equation \x \ then becomes
\eqn\z{i{\partial \psi_{\rm NR}\over \partial q^0}=\left[
	{({\bf k}+e{\bf A})^2\over 2m}+{es\over 2m^2}{\bf k\cdot F}
	-{es\over m}F_0 ~+eA_0\right]\psi_{\rm NR}+
{}~{\cal O}(F^2,\partial F, {1\over
	m^2},\cdots).}
It should be noted that, in our weak field approximation, we have
$t\equiv x^0=q^0+~{\cal O}(F^2)$ from equation \t.
Thus $k^0$ is indeed the time-translation
operator and so $H=i{\partial\over \partial t}=
i{\partial\over \partial q^0}+~{\cal O}(F^2, \partial F, \cdots)$ acting on
$\psi_{\rm NR}$.

Since $F^0$ is the magnetic field, we see from equation \z \ that the
anyon has a magnetic moment $\mu$ given by
\eqn\aa{\mu\equiv{e\over 2m}g s={es\over m}.}
In other words, we see again that the gyromagnetic ratio $g$ is 2. The term
${es\over 2 m^2}{\bf k\cdot F}$ contains the spin-orbit coupling.
Given $\mu$, one can derive the spin-orbit interaction by
Lorentz transformation of the electromagnetic fields, properly
taking account of Thomas precession. Needless to say, the spin-orbit
interaction in equation \z \ agrees with such a derivation as well.
(We have not considered quantized electromagnetic fields; these
results hold in the
limit of neglecting radiative corrections.)

Some remarks are in order. The terms we have neglected in our
expansions involve higher powers of the field and its derivatives.
It is clear that such terms cannot modify our results for
the magnetic moment and spin-orbit interaction. Secondly, the
commutation rules in equation \r \ have many solutions which
are unitary transforms of our solutions \s. If one uses one of these
other solutions, one should take account of the fact that the
Hamiltonian is not simply $k^0$ ($q^0$ is no longer representing
time) but the corresponding unitary transform. We kept $x^0\approx
q^0$ so that $k^0$ is indeed the time-translation operator.

Our analysis is clearly reminiscent of the Foldy-Wouthuysen (FW)
transformation of the Dirac theory. In a sense, we are
constructing the FW-transformation for anyons but without starting
from the covariant wave equation. The $x$-operators which do not
commute are similar to the noncommuting position operators of
the Dirac theory; the $q$'s are like the mean position operators.
As with the FW-transformation, once we have external fields,
our analysis can be done only in an expansion in powers and
derivatives of the field $^{10}$.
\vskip .2in
\leftline{\bf Note added: }
While this paper was being typed, we received a preprint `Electromagnetic
Interactions of Anyons in Nonrelativistic Quantum Field Theory,' by
J.L.Cort\'es, J.Gamboa and L.Vel\'azquez, Zaragoza-Orsay preprint,
hep-th number 9211106,
where anyons are obtained by coupling Dirac fields to a $U(1)$
Chern-Simons gauge field. The authors argue that a nonrelativistic
expansion yields $g=2$ for this model.
\vskip .2in
\noindent
{\bf References}
\vskip .1in
\item {1.} E.Merzbacher, {\it Am.J.Phys.} {\bf 30}, 237 (1960);
J.Leinaas and J.Myrheim, {\it Nuovo Cimento}, {\bf 37}, 1 (1977);
G.Goldin, R.Menikoff and D.Sharp, {\it J.Math.Phys.}, {\bf 21}, 650
(1980); {\bf 22}, 1664 (1981);
F.Wilczek, {\it Phys.Rev.Lett.}, {\bf 49}, 957 (1982); F.Wilczek and
A.Zee, {\it Phys.Rev.Lett.}, {\bf 51}, 2250 (1983); M.Bowick, D.Karabali
and L.C.R.Wijewardhana, {\it Nucl.Phys.}, {\bf B271}, 417 (1986); for a recent
review, see F.Wilczek, {\it Fractional Statistics and Anyon
Superconductivity} (World Scientific, Singapore, 1990).
\vskip .1in
\item {2.} R.B.Laughlin, {\it Phys.Rev.Lett.}, {\bf 50}, 1395 (1983);
for a recent review, see articles in {\it Physics and Mathematics of
Anyons}, Proceedings of the TCSUH Workshop, Houston, 1991, S.S.Chern,
C.W.Chu and C.S.Ting (eds.) (World Scientific, Singapore, 1991).
\vskip .1in
\item {3.} C.R.Hagen, {\it Ann.Phys.(NY)}, {\bf 157}, 342 (1984); {\it
Phys.Rev.}, {\bf D31}, 848 (1985); {\bf 31}, 2135 (1985); D.Arovas,
J.Schrieffer, F.Wilczek and A.Zee, {\it Nucl.Phys.}, {\bf B251}, 117
(1985); for a recent review, see R.Jackiw, in {\it Physics, Geometry and
Topology}, Proceedings of the NATO ASI, Banff, 1989, H.C.Lee (ed.)
(Plenum Press, New York, 1990).
\vskip .1in
\item {4.} B. Binegar, {\it J.Math.Phys.}, {\bf 23}, 1511 (1982).
\vskip .1in
\item {5.} A.P.Balachandran, G.Marmo, B.Skagerstam and A.Stern,
{\it Gauge Symmetries and Fibre Bundles} (Springer, Berlin, 1983) and
references therein;
P.Gerbert, {\it Nucl.Phys.}, {\bf B346}, 440 (1990); S.Forte and
T.Jolicoeur, {\it Nucl.Phys.}, {\bf B350}, 589 (1991).
\vskip .1in
\item {6.} R.Jackiw and V.P.Nair, {\it Phys.Rev.}, {\bf D43},
1933 (1991).
\vskip .1in
\item {7.} M.S.Plyushchay, {\it Phys.Lett.}, {\bf B248}, 107 (1990);
D.Shon and S.Khlebnikov, {\it JETP Lett.}, {\bf 51}, 611 (1990);
D.Volkov, D.Sorokin and V.Tkach, in {\it Problems in Modern Quantum Field
Theory}, A.Belavin, A.Klimyk and A.Zamolodchikov (eds.) (Springer,
Berlin, 1989).
\vskip .1in
\item {8.} See for example, V.I.Arnol'd, {\it Mathematical Methods of
Classical Mechanics}, (Springer-Verlag, New York, 1989); V.Guillemin and
S.Sternberg, {\it Symplectic Techniques in Physics} (Cambridge
University Press, Cambridge, 1990).
\vskip .1in
\item {9.} See for example, J.D.Jackson, {\it Classical
Electrodynamics} (John Wiley and Sons, Inc. 1975).
\vskip .1in
\item {10.} See for example, C.Itzykson and J-B.Zuber, {\it Quantum Field
Theory} (McGraw Hill, New York, 1980).
\end